\documentstyle[aps,prl,preprint]{revtex}
\begin{document}
\draft
\title{Electron spin relaxation by nuclei in semiconductor quantum dots}
\author{I. A. Merkulov}
\address{Ioffe Institute, Russian Academy of Science, St. Petersburg, Russia }
\author{Al. L. Efros and M. Rosen}
\address{ Naval Research Laboratory, Nanostructure Optics Section,
Washington DC 20375}
\maketitle
\begin{abstract}
We have studied theoretically the electron spin relaxation  in semiconductor quantum dots via interaction with nuclear spins. The relaxation is shown to be determined by three processes: (i) -- the precession of the electron spin
in the hyperfine field of the frozen fluctuation of the nuclear spins; (ii) -- the precession of the nuclear spins in the hyperfine field of the electron; and (iii) -- the precession of the nuclear spin in the dipole field of its nuclear neighbors.
In external magnetic fields the relaxation of electron spins directed along the magnetic field is  suppressed. Electron spins directed transverse to the magnetic field relax completely in a time on the order  of the precession period of its spin in the field of the frozen fluctuation of the nuclear spins. Comparison with experiment shows that the hyperfine interaction with nuclei  may be the dominant mechanism of electron spin relaxation  in quantum dots.
\end{abstract}


\narrowtext
\section{Introduction}
The long  electron spin dephasing time (hundred nanoseconds) recently  reported \cite{Awshalom}
suggests using the spin of an electron localized in a quantum dot (QD) as the realization of a quantum bit, and electron doped quantum dots as the basic material for implementing  a solid state based quantum computer \cite{LossDi}.  Room temperature operation, which is usually a primary requirement in microelectronics, is of secondary importance for
quantum computers relative to that of finding a material for their physical realization. Rather, operating now at low temperature, where the localized electron has a long spin coherence time, is an essential condition for quantum computing and spin memory
storage.  The dominant  electron spin relaxation mechanism in bulk is connected with the spin-orbit  interaction of carriers (see Ref.\cite{OO}), but  is strongly suppressed for localized carriers \cite{Khaetzkii}.  Spin relaxation due to the electron hole exchange interaction plays an important  role during the time of nonequilibrium carrier relaxation \cite{Dzhioev,Nickolaus,Maialle}, but does not effect localized electron spin relaxation afterword. Dephasing of the electron in QD ground
states via two phonon real transitions to, or virtual transitions through, excited states (Urbach or Raman two phonon processes \cite{vanflek})  is also suppressed at low temperatures \cite{Takagahara}. As a result electron spin relaxation via interaction with nuclei becomes the dominant relaxation mechanism  for  localized electrons at low temperatures.

The interaction of localized electrons  with nuclei was studied early on for electrons localized at donors in bulk (see for example \cite{Lampel,DP}).  There, the electron interacts with a large number of nuclei and feels the hyperfine magnetic field  of the nuclei located in the region where electron is localized; this is also true for QD localized electrons. However, the correlation time of the electron-nucleus interaction of donor localized electrons is limited by the time of shallow donor ionization, and
 tunnel jumps between the donors, $\tau_c$. Usually this time is much shorter  than the period of the electron spin precession, $\omega_f^{-1}$, in the  hyperfine field of the nuclei. For quantum computation and spin storage, however, it is important to have a large value of $\tau_c$, and we are interested in the opposite limit: $\omega_f\tau_c\gg 1$. This is realized   in QDs at sufficiently low temperatures.

In this paper we consider the electron spin relaxation via its interaction with the spins of the nuclei in QDs in the absence and the presence of an external magnetic field.
The process is facilitated by the disparity of  the characteristic time scales of the three processes that determine the relaxation: the period of the electron precession in the frozen fluctuation of the hyperfine field of the nuclei,  the period of the nuclear spin precession in the hyperfine field of the electron, and the nuclear spin relaxation time in the dipole-dipole field of its nuclear neighbors.  Estimates of these time scales can be made for the case of GaAs, whose  hyperfine constants  are well k
own. For QDs containing 10$^5$ nuclei they are found to be: $\sim 1$\,ns, $\sim 1$\,$\mu$s and $\sim 100$\,$\mu$s,  respectively. Therefore, as a first step, we can describe the electron spin
relaxation as a precession in the quasi-stationary frozen fluctuation of the hyperfine  field of the nuclear spins. We can than examine the additional electron spin relaxation arising from the slowly  varying nuclear hyperfine fields.  The nuclear dipole-dipole interaction does not conserve the total nuclear spin; the third time scale provides a natural limit on the coherence of the electron-nuclei spin system.  However many other  relaxation mechanisms are important at this long time scale, therefore and we will not consider the effect of the nuclear dipole-dipole interactions on the electron spin relaxation. A spin dephasing  time ($T_2^*$) calculated for an ensemble of QDs is in a good agreement with avaliable experimental data.

The paper is organized as  follows: The hyperfine interaction of an electron with the nuclei and the hyperfine nuclear magnetic fields acting on the electron  are considered in section II. The electron spin relaxation times in the absence and in the presence of an external magnetic field are calculated in sections III and IV respectively. In section V we compare our theoretical results with experimental data.

\section{Hyperfine interaction of a localized electron with nuclei. The Frozen Fluctuation of  the nuclear hyperfine field in quantum dots.}

The electron spin relaxation  due to the nuclear spins is determined by their hyperfine Fermi contact interaction. The Hamiltonian of this interaction can be written \cite{Abragam}:
\begin{equation}
\hat{H}_{\rm cont}={16\pi\over 3}\mu_B\sum_j {\mu_j\over I_j}(\mbox{\boldmath $\hat{S}\cdot \hat{I}$}_j)\delta(\mbox{\boldmath $r$}-\mbox{\boldmath $R$}_j) ~,
\end{equation}
where $\mu_B$ is the Bohr magneton,  {\boldmath $\hat{S}$} and {\boldmath $r$} are the spin and position of the electron, $\mu_j$,  {\boldmath $\hat{I}$}$_j$ and {\boldmath $R$}$_j$ are the magnetic moment, spin and position of the $j$-th nucleus, and the sum goes over all the nuclei in the lattice.   For localized electrons the  distance between their energy levels is much larger than the energy of the hyperfine interaction with the nuclei.  As a result this spin-spin interaction can be described by the Hamiltonian:
\begin{equation}
\hat{H}_{\rm hf}= {v_0\over 2}\sum_jA^j|\psi(\mbox{\boldmath $R$}_j)|^2\left( \hat{I}_z^j\hat{\sigma}_z+\hat{I}^j_x\hat{\sigma}_x+\hat{I}^j_y\hat{\sigma}_y\right)~,
\label{interaction}
\end{equation}
which is obtained from first order perturbation theory. Here $v_0$ is the volume of the unit cell,
$\psi(\mbox{\boldmath $R$}_j)$ is the electron envelope wave function at the $j$-th nucleus, $\hat{I}_{\alpha}$ and $\hat{\sigma}_{\alpha}$ are the  spin projections on the coordinate axes $\alpha=x,y,z$, and $A^j=(16\pi\mu_B\mu_j/3I^j)|u_c(\mbox{\boldmath $R$}_j)|^2$, where $u_c(\mbox{\boldmath $R$}_j)$ is the electron Bloch function at the nucleus. In GaAs, the sum of $A^j$ over all the nuclei in the unit cell
$A=\sum_j A^j\approx 90$\,$\mu$eV \cite{Paget}. We can neglect the interaction of holes with the nuclei because the hole Bloch functions vanish at the nuclear positions.  In addition, we neglect, for now, the nuclear dipole-dipole  interactions, which do not conserve the total spin of the  electron-nuclear system. They become important only at times longer than 10$^{-4}$\,s.

 The effective nuclear hyperfine magnetic field,  $\mbox{\boldmath $B$}_N$, acting on a localized electron spin can be obtained
 from Eq.\ref{interaction} taking the expectation of the Hamiltonian $\hat{H}_{\rm hf}$
over the ensemble of nuclear wave functions. It is the sum of contributions from a large
number of nuclei:
\begin{equation}
\mbox{\boldmath $B$}_N = {\nu_0\over \mu_Bg_e}\langle \sum_jA^j|\psi(\mbox{\boldmath $R$}_j)|^2\hat{\mbox{\boldmath $I$}}^j \rangle_{N}
\label{field}
\end{equation}
where $\langle ...\rangle_N$ denotes a quantum mechanical average over the ensemble of nuclear wave functions and $g_e$ is the electron $g$-factor. We should note that the precession frequency of the electron in the hyperfine field of all the nuclei is much greater then the precession frequency of a nucleus in the hyperfine field of the electron. That is,
the electron sees a snapshot of the "frozen fluctuation" of the nuclear field.  The magnitude and direction of this
field are randomly distributed, and described by a Gaussian probability density distribution function:
\begin{equation}
W(\mbox{\boldmath $B$}_N)= {1\over \pi^{3/2}\Delta_B^3} \exp\left[-{(\mbox{\boldmath $B$}_N)^2\over \Delta_B^2}\right]~,
\label{distrB}
\end{equation}
where $\Delta_B$ is the dispersion of the nuclear hyperfine field distribution,
\begin{equation}
\Delta_B^2={2\over 3}\langle (\mbox{\boldmath $B$}_N)^2\rangle={2\over 3}\sum_jI^j(I^j+1)(a_j)^2~,
\end{equation}
where
\begin{equation}
a_j=(v_0/\mu_Bg_e)A^j|\psi(\mbox{\boldmath $R$}_j)|^2
\label{aj}
\end{equation}
is the  magnetic field of a single nuclear spin acting on the electron,
and we assumed that the nuclear spin directions are independent of each other.
All nuclei in GaAs have the same spin $I^j=I=3/2$. Replacing the sum over unit cells by an integration we obtain:
 \begin{equation}
\Delta_B^2={2I(I+1)\over 3}{\sum_j (A^j)^2\over (\mu_Bg_e)^2}{v_0\over V_{\rm L}}={16I(I+1)\over 3N_L}{\sum_j (A^j)^2\over (\mu_Bg_e)^2}~,
\label{dispersion1}
\end{equation}
where the  sum in this equation goes over only those nuclei in a unit cell,
\begin{equation}
 V_{\rm L}=(\int d^3r\psi^4(\mbox{\boldmath $r$}))^{-1}~{\rm and}~N_L=8V_L/v_0
\end{equation}
is the number of nuclei in the volume, $V_{\rm L}$, that effectively determine the electron precession frequency. In GaAs, the sum $\sum_j (A^j)^2\approx 1.2\cdot10^{-3}$\,meV$^2$. The volume, $V_{\rm L}$, is on the order of  the volume of the electron localization.

\section{Electron spin relaxation in zero external magnetic field}
\subsection {Electron spin dephasing in the frozen fluctuations of the nuclear field}

Let us consider an ensemble of identical QDs in which we simultaneously (at time $t=0$) create electrons all
having the same spin orientation (They can be created, for example,  by circularly polarized light). The nuclear spins in the QDs of this ensemble are randomly oriented, the nuclear hyperfine fields in the dots differ from one another and, therefore,  have a  different effect on the initial electron spin, $\mbox{\boldmath $S$}_0$, in each dot. We will consider the time dependence of the ensemble average electron spin relaxation for times small relative to the period of the nuclear precession in the hyperfine
 field of the electron.  Each electron spin will be moving in the frozen fluctuation of the nuclear hyperfine magnetic field,
$\mbox{\boldmath $B$}_N$, (see Eq.\ref{field}) in its own QD.  These fields, however, are randomly distributed among the dots of the ensemble.  Therefore, even though each electron spin will precess in a coherent fashion in the frozen hyperfine field of its own dot, the ensemble average spin polarization will decrease.

The equation of  motion of the spin ${\boldmath S}$ in a fixed magnetic field, ${\boldmath B}$, is given by:
\begin{eqnarray}
\mbox{\boldmath $S$}(t)&=&(\mbox{\boldmath $S$}_0\cdot \mbox{\boldmath $n$})\mbox{\boldmath $n$}+\{\mbox{\boldmath $S$}_0-(\mbox{\boldmath $S$}_0\cdot \mbox{\boldmath $n$})\mbox{\boldmath $n$}\}\cos\omega t\nonumber \\
&+&[\{\mbox{\boldmath $S$}_0
-(\mbox{\boldmath $S$}_0\cdot \mbox{\boldmath $n$})\mbox{\boldmath $n$}\}\times \mbox{\boldmath $n$}]\sin\omega t ~,
\label{smotion}
\end{eqnarray}
where $\mbox{\boldmath $S$}_0$ is the initial spin, $\mbox{\boldmath $n$}=\mbox{\boldmath $B$}/B$ is a unit vector in the direction of the magnetic field, and $\omega=\mu_Bg_eB/\hbar$ is the Larmor frequency of the electron  precession in this field.  The equation also describes  the coherent electron spin precession in a single QD due to the  magnetic field, $\mbox{\boldmath $B$}_N$, of the frozen fluctuation of the nuclei ($\mbox{\boldmath $n$}=\mbox{\boldmath $B$}_N/B_N$, and $\omega=\mu_Bg_eB_N/\hbar$).
 Averaging Eq.(\ref{smotion}) over the magnetic field distribution of Eq.(\ref{distrB}),
we obtain the  time dependence of the ensemble averaged electron spin polarization:
\begin{equation}
\langle \mbox{\boldmath $S$}(t)\rangle={\mbox{\boldmath $S$}_0\over 3}\left\{1+2\left[1-2\left({t\over T_{\Delta}}\right)^2\right]\exp\left[-\left({t\over T_{\Delta}}\right)^2\right]\right\}~.
\label{zerofield}
\end{equation}
The same time dependence describes the electron spin polarization of a single quantum  dot  averaged over a large number of measurements. Here
\begin{equation}
T_{\Delta}={\hbar\over \mu_Bg_e\Delta_B}=\hbar \sqrt{{3N_{\rm L}\over 16\sum_j I^j(I^j+1)(A^j)^2}}
\end{equation}
 is the ensemble dephasing time which arises from the random electron precession frequencies in the randomly distributed frozen  fluctuation of the nuclear hyperfine field in the dots. This time is on the order of $1$\,ns  for GaAs quantum dots with $10^5$ nuclei.  The spin dephasing time is proportional to $\sqrt{V_{\rm L}}$. One can see that the average electron polarization relaxes to 10\% of its original value after a time equal to the dephasing time and then increases to a steady state value of 33\% of
its initial polarization.

\subsection{Electron spin dephasing as a result of variations of the nuclear field direction}

A localized electron interacts with a large number of nuclei, $N_L\gg 1$. The  interaction of the electron spin with a single nucleus is $\sqrt{N_L}$ times weaker than its interaction with the effective magnetic field of the frozen nuclear fluctuation. Changing the  direction of a single nuclear spin only weakly perturbs the electron spin motion. The precession of the electron in the macroscopic fluctuation of the nuclear spins is $\sqrt{N_L}$ times faster than the precession of a nucleus in the hyperfine field of an electron, i.e., the nuclear precession period $T_N\sim T_\Delta\sqrt{N_L}$. Since
an electron in a QD precesses so rapidly around the nuclear magnetic field
 $\mbox{\boldmath $B$}_N$,  the nuclei only see the long time average of the hyperfine field of the electron, which is directed along $\mbox{\boldmath $B$}_N$. Note, that components of the nuclear field perpendicular to $\mbox{\boldmath $B$}_N$ cancel each other out. Now, however, each nucleus, in turn, precesses about this direction with a different precession rate that is proportional to the square of the electron wave function at their respective nuclear positions. This variation in the precession rates leads to a non-vanishing slow time varying change in the frozen fluctuation of the hyperfine magnetic field of the nuclei, $\Delta \mbox
{\boldmath $B$}_N\sim \sum_j\left(A^j|\psi(\mbox{\boldmath $R$}_j)|^2\right)^2\langle\hat{\mbox{\boldmath $I$}}^j\times \mbox{\boldmath $B$}_N\rangle_{N}$, that is perpendicular to $\mbox{\boldmath $B$}_N$.  These random changes in the nuclear magnetic field result in an additional relaxation of the  electron spin polarization.

Equation (\ref{zerofield}) describes the time dependence of the electron spin relaxation in the frozen fluctuation of the nuclear hyperfine fields for times much less than $T_N$.  To include the effect of the time dependent changes in the nuclear fields, we need to examine the ensemble average of Eq.(\ref{smotion}) at a much later time than is considered in Eq.(\ref{zerofield}). Consider the ensemble average of Eq.(\ref{smotion}), but now at times for which the nuclear fields are time dependent.  In addition, we average this quantity over a time interval large compared to the period of the electron precession but small compared to the time at which the ensemble average is taken. This leads to:
 \begin{equation}
\langle \mbox{\boldmath $S$(t)}\rangle= \langle \mbox{\boldmath $n$}(t)(\mbox{\boldmath $n$}(t)\cdot \mbox{\boldmath $S$}(t))\rangle=
\left\langle{\mbox{\boldmath $B$}_N(t)(\mbox{\boldmath $B$}_N(t)\cdot \mbox{\boldmath $S$}(t))\over B_N^2(t)}\right\rangle ~~.
\end{equation}
In this equation $(\mbox{\boldmath $B$}_N(t)\cdot \mbox{\boldmath $S$}(t))$ is the energy of the electron-nucleus spin system, which does not depend on time and is equal to $(\mbox{\boldmath $B$}_N(0)\cdot \mbox{\boldmath $S$}(0))$. Nor does $B_N^2$  change its value, because $d\mbox{\boldmath $B$}_N/dt \perp \mbox{\boldmath $B$}_N$.   As a result,   \begin{equation}
\langle \mbox{\boldmath $S$}(t)\rangle= \left\langle{\mbox{\boldmath $B$}_N(t)(\mbox{\boldmath $B$}_N(0)\cdot \mbox{\boldmath $S$}(0))\over B_N^2(0)}\right\rangle~~.
\label{st}
\end{equation}
 The rotational symmetry of the Hamiltonian Eq.{\ref{interaction} leads to  $\langle\mbox{\boldmath $B$}_N(t)_\alpha
\mbox{\boldmath $B$}_N(0)_\beta \rangle= \delta_{\alpha,\beta}\langle{\bf B}_N(t)_x {\bf B}_N(0)_x \rangle=(\delta_{\alpha,\beta}/3)\langle(\mbox{\boldmath $B$}_N(t)\cdot\mbox{\boldmath $B$}_N(0))\rangle$. This allows us to write Eq.\ref{st} as
 \begin{equation}
\langle \mbox{\boldmath $S$}(t)\rangle= \left\langle{(\mbox{\boldmath $B$}_N(t)\cdot \mbox{\boldmath $B$}_N(0))\over B_N^2(0)}\right\rangle{\mbox{\boldmath $S$}(0)\over 3}~~.
\label{correlator}
\end{equation}
We see again that $ \langle \mbox{\boldmath $S$}(t)\rangle$ is directed along $\mbox{\boldmath $S$}(0)$, which is the only physically defined direction in the ensemble.   This equation describes the time dependence of the electron spin relaxation at times large compared to $T_\Delta$ and is limited only by the characteristic time scale of the dipole-dipole interactions. One can see that at times $T_\Delta<t<T_N$  Eq.(\ref{correlator}) gives the same result as Eq.(\ref{zerofield}).

The variation of nuclear magnetic field direction in the time dependent correlation, in Eq.(\ref{correlator}), is limited by the conservation of the total spin angular momentum of the electron-nucleus spin system ,  $\hat{\mbox{\boldmath $F$}}=\hat{\mbox{\boldmath $S$}}+\sum_j\hat{\mbox{\boldmath $I$}}^j$.  The conservation of $\hat{\mbox{\boldmath $F$}}$ at times shorter than the nuclear dipole-dipole relaxation time follows from the fact that $\hat{\mbox{\boldmath $F$}}$ commutes with the Hamiltonian of t
e electron-nucleus spin system Eq.(\ref{interaction}).
The total nuclear spin is, then,  effectively conserved since: $\mbox{\boldmath $I$}_\Sigma=\sum_j \mbox{\boldmath $I$}_j=\mbox{\boldmath $F$}
-\mbox{\boldmath $S$}\approx \mbox{\boldmath $F$}$ because $F \gg S$ \cite{diffusion}. The latter follows from the fact that the dispersion of  $\mbox{\boldmath $I$}_\Sigma$ increases with the number of nuclei in each QD and the average value of $|\mbox{\boldmath $I$}_\Sigma|\gg 1$.

If the electron wave function were constant in the  localization region and zero outside
(the so called "box model" \cite{Ryabchenko})  the nuclear magnetic field, $\mbox{\boldmath $B$}_N$, would be proportional to the total nuclei spin, $\mbox{\boldmath $I$}_\Sigma$, and would also be conserved. From Eq.(\ref{correlator})
the nuclear spin precession would not then lead to any additional spin relaxation in this model, and  $\langle \mbox{\boldmath $S$}(t)\rangle=\mbox{\boldmath $S$}_0/3$.
However, in real QDs the amplitude of the electron wave function at the nuclei in the localization region does depend on their  position.  The nuclear field is not uniquely determined by  the value and direction of the  total nuclear spin; a
distribution of values of $\mbox{\boldmath $B$}_N$ are possible for the same value of $\mbox{\boldmath $I$}_\Sigma$.  The total nuclear spin, $\mbox{\boldmath $I$}_\Sigma$, can be distributed in different ways among the nuclei as a result of their interaction with the electron.

          The determination  of the correlation, Eq.(\ref{correlator}) at all times is beyond the scope of this paper. We will evaluate it  only in the limit $t\gg T_N$, where the $\mbox{\boldmath $B$}_N(t)$ are randomly distributed.   The joint distribution function of the nuclear fields and total nuclear spin  can be written as the product of the total nuclear spin
distribution function and the conditional probability distribution of nuclear fields given a certain value of the total nuclear spin:
\begin{equation}
W(\mbox{\boldmath $B$}_N,\mbox{\boldmath $I$}_\Sigma)= W_I(\mbox{\boldmath $I$}_\Sigma)w(\mbox{\boldmath $B$}_N|\mbox{\boldmath $I$}_\Sigma)~.
\label{distr2}
\end{equation}
Both of these latter distributions have a Gaussian form \cite{LL}:
\begin{equation}
W_I(\mbox{\boldmath $I$}_\Sigma)={1\over \pi^{3/2}\Delta_I^3} \exp\left[-{( \mbox{\boldmath $I$}_\Sigma)^2\over \Delta_I^2}\right]~,
\label{distrI}
\end{equation}
and
\begin{equation}
w(\mbox{\boldmath $B$}_N|\mbox{\boldmath $I$}_\Sigma)={1\over \pi^{3/2}\Delta_B^{'3}} \exp\left[-{(\mbox{\boldmath $B$}_N-\langle a\rangle
\mbox{\boldmath $I$}_\Sigma)^2\over \Delta_B^{'2}}\right]~,
\label{distrBI}
\end{equation}
where $\Delta_I^2=2NI(I+1)/3$ is  the dispersion
 of the total nuclear spin distribution, and $N$ is the number of nuclei which are in statistical equilibrium with the localized electron (see Appendix). This number is  approximately equal to the number of nuclei in the dot.  The conditional distribution of the random hyperfine fields, given a value  $\mbox{\boldmath $I$}_\Sigma$ for the total nuclear spin, is shifted from zero to the field due to  the weighed total nuclei spin, $\langle a\rangle\mbox{\boldmath $I$}_\Sigma$, where
  $\langle a\rangle=(\sum_j^N a_j)/N=A/(\mu_Bg_eN)$. The dispersion of this distribution, $\Delta_B^{'2}= \Delta_B^2 -\langle a\rangle^2\Delta_I^2 $,  is determined by integrating Eq.\ref{distrBI} with  distribution Eq.(\ref{distrI}) over all $\mbox{\boldmath $I$}_\Sigma $  and comparing it with Eq.(\ref{distrB}).  Clearly $\Delta_B^{'2}<\Delta_B^2$, because  the latter also includes the dispersion of the random distribution of the total  nuclear spin.

In general, to calculate the correlation between the direction of the hyperfine nuclear field  at two times, $\mbox{\boldmath $n$}_0=\mbox{\boldmath $B$}_N(0)/B_N(0)$, and  $\mbox{\boldmath $n$}(t)=\mbox{\boldmath $B$}_N(t)/B_N(t)$, (see Eq.(\ref{correlator})),
we have to find the angular distribution of the magnetic field $\mbox{\boldmath $B$}_{N}(t)$ given a  certain value both of the total nuclear spin  $\mbox{\boldmath $I$}$ and the  magnetic field magnitude, $B_N=|\mbox{\boldmath $B$}_{N}|$. This conditional distribution function is written:
\begin{equation}
F(\mbox{\boldmath $n$}|B_N,\mbox{\boldmath $I$})={\exp\left[-(B_N\mbox{\boldmath $n$}-\langle a\rangle \mbox{\boldmath $I$})^2/\Delta^{'2}_B\right]\over
\int d\mbox{\boldmath $\Omega$}(\mbox{\boldmath $n$})\exp\left[-(B_N\mbox{\boldmath $n$}-\langle a\rangle \mbox{\boldmath $I$})^2/ \Delta^{'2}_B\right]}~,
\label{distrn}
\end{equation}
Using  distributions of Eqs.(\ref{distrB},\ref{distrI},\ref{distrn}) we find  the correlation Eq.(\ref{correlator})
\begin{eqnarray}
&&\gamma=\langle(\mbox{\boldmath $n$}\cdot\mbox{\boldmath $n$}_0)\rangle\nonumber\\
&=&
\int(\mbox{\boldmath $n$}\cdot\mbox{\boldmath $n$}_0)F(\mbox{\boldmath $n$}|B_N,\mbox{\boldmath $I$})w(\mbox{\boldmath $B$}_N|\mbox{\boldmath $I$})W_I(\mbox{\boldmath $I$})d^3B_Nd\mbox{\boldmath $\Omega$}(\mbox{\boldmath $n$})d^3I~.\nonumber\\
\end{eqnarray}
This gives us the long time ($t\gg T_N$) electron spin polarization (Eq.(\ref{correlator}).
 Straightforward calculation gives:
\begin{eqnarray}
\gamma(x)&=&{2\over \pi x^3}\int_0^\infty dy\int_0^{\infty}dz{[2y~\cosh(2y)-\sinh(2y)]^2\over y z~\sinh(2y) }\nonumber\\
&\times&\exp\left[-\left(1+{1\over x^2}\right)z^2-\left({y \over z}\right)^2\right]~,
\label{gamma}
\end{eqnarray}
where
\begin{equation}
x={\langle a \rangle ^2\Delta_I^2\over \Delta_B^{'2}}
={\langle a \rangle ^2\over \langle a^2 \rangle -\langle a \rangle ^2}
\end{equation}
is the relative  dispersion of the hyperfine magnetic field of the localized electron acting on nuclei and $\langle a^2 \rangle=\Delta_B^{2}/ \Delta_I^2$.  If all the nuclei in the unit cell have the same hyperfine constant $x=N_L/(N-N_L)$.

In Fig. 1. we show $\gamma $ as function of  $\langle a^2 \rangle/\langle a \rangle ^2\approx N_L/N$.  In the limit of large $x$, where  the number of nuclei contributing to  the field, $\mbox{\boldmath $B$}_N$, is close to the number of nuclei whose total nuclear spin is conserved, $\gamma (x)\approx 1$. This case is realized in the "box model"  and also at short times $t< T_N$. The number of nuclei that are included in the electron-nucleus spin system of the Hamiltonian Eq.(2), and whose  total spin is conserved,  increases with time, which in turn decreases $\gamma$. However this only holds for $t< T_{d-d}$ the dipole-dipole nuclear spin relaxation time in the dipole field of its neighbors.
For  spherical QDs with 10$^5$ nuclei confined in an infinite potential barrier (note that $N_L\approx N_{\rm tot}/2.8$, where $N_{\rm tot}$ is the total number of nuclei in the QD), our  approach is valid for the time interval $10T_N<t<100T_N$ (see Appendix). The calculation in this model shows that the number of nuclei that need be included  at $t\sim T_{d-d}$ is $\sim 2.1N_L$ ($\gamma\approx 0.4$) and does not reach its maximum value $N\approx 2.8N_L$.

It is important to note that non vanishing average spin polarization at time $t\gg T_N$ ($\gamma\neq 0$) means that there is a significant probability that an ensemble of  electron-nucleus spin systems retains its initial spin state.

\section{Spin decoherence in  a strong external magnetic field.}
A strong external magnetic field, $B$, ($B\gg B_N$) significantly changes the process of electron spin
relaxation. In  this large field the Zeeman splitting of the electron spin levels is larger than their inhomogeneous broadening in the hyperfine nuclear magnetic field. The total magnetic field acting on the electron is now effectively directed along the external magnetic field. The nuclear hyperfine fields only perturb the precession frequency of the electron spin  about the external magnetic field direction.

Consider, now, the effect of a strong external magnetic field on the electron spin polarization.  The  motion  of the spin in the total magnetic field is again described by Eq. (\ref{smotion}) where, now,  $\mbox{\boldmath $n$}=(\mbox{\boldmath $B$}+\mbox{\boldmath $B$}_N)/|\mbox{\boldmath $B$}+\mbox{\boldmath $B$}_N|$.
Averaging Eq. (\ref{smotion}) over the ensemble, using the  distribution  of nuclear magnetic fields in Eq.(\ref{distrB}), we obtain:
\begin{eqnarray}
\langle \mbox{\boldmath $S$}(t)\rangle &=&R_{\|}(t)
(\mbox{\boldmath $S$}_0\cdot\mbox{\boldmath $b$})\mbox{\boldmath $b$}+R_\bot^0(t)[\mbox{\boldmath $S$}_0-(\mbox{\boldmath $S$}_0\cdot\mbox{\boldmath $b$})\mbox{\boldmath $b$}]\nonumber\\
&+&R_{\bot}^1(t)[(\mbox{\boldmath $S$}_0-(\mbox{\boldmath $S$}_0\cdot\mbox{\boldmath $b$})\mbox{\boldmath $b$})\times \mbox{\boldmath $b$}]~,
\label{stb}
\end{eqnarray}
where
\begin{eqnarray}
R_\|(t)&=&R_\|^\infty+\Delta R_{\|}(t)~, \nonumber\\
R_\bot^0(t)&=&R^\infty_\bot+\Delta R_{\bot}^0(t)~,\nonumber
\end{eqnarray}
and $\mbox{\boldmath $b$}=\mbox{\boldmath $B$}/B$ is a unit vector along the external magnetic field,
\begin{equation}
R^\infty_\|=\left\langle {[B+(\mbox{\boldmath $B$}_N\cdot\mbox{\boldmath $b$})]^2\over |\mbox{\boldmath $B$}+\mbox{\boldmath $B$}_N|^2}\right\rangle=1-2 R^\infty_\bot~,\label{23}
\end{equation}
and
\begin{eqnarray}
 R^\infty_\bot &=&{1\over 2}\left\langle {B_N^2-(\mbox{\boldmath $B$}_N\cdot\mbox{\boldmath $b$})^2\over |\mbox{\boldmath $B$}+\mbox{\boldmath $B$}_L|^2}\right\rangle\nonumber\\
&=&{1\over 2\sqrt{\pi}}\int_{-\infty}^\infty dz\int_0^\infty dy{y\exp(-z^2-y)\over (\beta+z)^2+y}
\label{Rinf}
\end{eqnarray}
is the value of $\langle {\bf S}(t)\rangle$ in the long time limit
$t\gg T_{\Delta}$, and $\beta= B/\Delta_B$. The time dependent components are given by
\begin{eqnarray}
\Delta R_\|(t)&=&{1\over \sqrt{\pi}}\int_{-\infty}^\infty dz\int_0^\infty dy{y\exp(-z^2-y)\over (\beta+z)^2+y}\nonumber\\
&\times&\cos\left[\sqrt{(\beta+z)^2+y}{t\over T_{\Delta}}\right]~,\nonumber\\
\Delta R_{\bot }^0(t)&=&E(t)-{\Delta R_\|(t)\over 2}~, \label{Rt}
\end{eqnarray}
where
\begin{equation}
E(t)=\exp\left[-\left({t\over 2T_{\Delta}}\right)^2\right]\left[\cos\left({\beta t\over T_{\Delta}}\right)-{t\over 2\beta T_{\Delta}}\sin\left({\beta t\over T_{\Delta}}\right)\right]~,\nonumber
\end{equation}
and
\begin{eqnarray}
 R_{\bot }^1(t)&=&{1\over \sqrt{\pi}}\int_{-\infty}^\infty dz\int_0^\infty dy{[(\beta+z)^3+zy]exp(-z^2-y)\over [(\beta+z)^2+y]^{3/2}}\nonumber\\
&\times& \sin\left[\sqrt{(\beta+z)^2+y}{t\over T_{\Delta}}\right]~.
\end{eqnarray}
Equation (\ref{stb}) simplifies considerably  for  strong magnetic fields. Calculation of the coefficients in
Eqs. (\ref{23},\ref{Rinf},\ref{Rt}) in the limit $\beta\gg 1$ gives:
\begin{eqnarray}
&&\langle\mbox{\boldmath $S$}(t)\rangle\approx \nonumber\\ &&\left\{1-{ 1-\cos\left(\omega_Bt\right)\over \beta^2} \exp\left[-\left({t\over 2T_\Delta}\right)^2\right]\right\}(\mbox{\boldmath $S$}_0\cdot \mbox{\boldmath $b$})\mbox{\boldmath $b$} \nonumber\\
&+&\left\{\left[\cos(\omega_Bt)+{1-\cos\left(\omega_Bt\right)\over 2\beta^2} \right][\mbox{\boldmath $S$}_0-(\mbox{\boldmath $S$}_0\cdot \mbox{\boldmath $b$})\mbox{\boldmath $b$}]\right. \nonumber\\
&+&\left.\sin(\omega_Bt)[(\mbox{\boldmath $S$}_0-(\mbox{\boldmath $S$}_0\cdot \mbox{\boldmath $b$})\mbox{\boldmath $b$})\times \mbox{\boldmath $b$}]\right\}\exp\left[-\left({t\over 2T_\Delta}\right)^2\right]~.
\label{strongfield}
\end{eqnarray}
One can see that in strong magnetic fields, $B\gg \Delta_B$ the component of spin along $\mbox{\boldmath $B$}$ is conserved, while its two transverse components precess with a frequency $\omega_B=\mu_Bg_eB/\hbar$, and decay as a result of
the inhomogeneous broadening of the levels in the random magnetic field of nuclei, respectively .  The dephasing arises from the dispersion of the nuclear field along the external magnetic field, which leads to an inhomogeneous dispersion of the electron precession frequency. The perpendicular components of the nuclear magnetic field change the direction of the precession axis by a small angle of $\sim B_N/B$ and   lead to a dephasing rate that is $(B_N/B)^2$ smaller than that due
to the dispersion
of the nuclear field along the external field. This result is consistent with the small ratio of the {\em lifetime broadening} to the {\em secular broadening} describing the transverse relaxation time $T_2$ in strong magnetic fields (see e.g. Ref. \cite{Slichter}).

 Figure 2 shows the time dependence of the various components of $\langle \mbox{\boldmath $S$}(t)\rangle$ which occur in Eq.(\ref{stb}). The both longitudinal and transverse components of the electron spin polarization  tend to a steady state value after several oscillations.  The number of oscillations grows  with increasing a magnetic field. Increasing the magnetic field also changes the steady state value of the longitudinal component $R_{\|}( \infty )$ (from 1/3 to 1) and
 the transverse component $R_{\bot}^{0}(\infty)$ (from 1/3 to 0), while  the steady state value of $R_{\bot}^{1}(\infty)$ is zero for all values of $B$.  In a strong magnetic field when the nuclear spin relaxation mechanism of the longitudinal spin polarization is suppressed phonons can again play an important role (see ref. \cite{nazarov}).

The important characteristic measured in steady state  experiments, such as Hanle effect measurements (e.g., see \cite{OO}), is the average electron polarization for its lifetime, $(1/\tau)\int \langle S(t)\rangle \exp(-t/\tau) dt$,  where $\tau$ is the lifetime of the localized electrons. Comparison with Eq.(\ref{stb}) shows that this average polarization is characterized by:
\begin{eqnarray}
\rho_\|(\beta,\tau)&=&{1\over \tau}\int_0^\infty R_\|(t)\exp(-t/\tau) dt~,\nonumber\\
\rho_{\bot}^{0,1}(\beta,\tau)&=&{1\over \tau}\int_0^\infty R_\bot^{0,1}(t)\exp(-t/\tau) dt~.
\end{eqnarray}
The dependence of these respective terms on the magnetic field is shown  in Fig. 3 for different values of the electron lifetime.

Equation (\ref{stb}), as is Eq.(\ref{zerofield}), is derived assuming a time independent frozen fluctuation of the nuclear field acting on the electrons. In zero external magnetic field, this no longer holds at longer times such that we must take into account  the nuclear spin precession in the inhomogeneous hyperfine field of the electron. Nuclear spins precessing at different rates about the average electron spin direction create a time dependent hyperfine field with components perpendicular to the original direction of the frozen fluctuation. As a result  the average electron spin projection follows the new direction of the slowly varying nuclei field.  The characteristic time of this slow process is determined by the dispersion  of the nuclear spin precession frequency in the inhomogeneous field of the electron $T_N^{-1}\approx (\mu_Bg_e/\hbar)\sqrt{\langle a^2\rangle -\langle a\rangle^2}\sim (\mu_Bg_e/\hbar)\sqrt{\langle a^2\rangle}$. This second regime of spin relaxation begins  when $t\geq T_N$.

In a strong external magnetic field $B\gg \Delta_B$, the average electron spin is directed along this strong field, independent of the nuclear hyperfine fields (${B}_N\ll B$).  Although the nuclei precess with different frequencies in the inhomogeneous electron field, the electron is effected only by the component of the nuclear field along the external field.  As a result  the  nuclear magnetic field acting on the electron spin is frozen for  times much longer than $T_N$. Thus a frozen fluctuation model of
the nuclear hyperfine field is valid when describing the dephasing dynamics of the electron spin polarization in an  ensemble of quantum dots in strong magnetic fields. As we mentioned above this consideration
is limited by a low enough temperature and the time scale of the nuclear dipole-dipole interaction.

In each QD, the motion of the electron spin in the hyperfine field of the frozen fluctuation of the nuclei is coherent. The dephasing is a result of inhomogeneous broadening  of the electron spin levels in the ensemble of quantum dots. This makes it possible to recover the transverse electron spin polarization using the spin echo technique \cite{Abragam}, which also can be used for quantum computation \cite{Chuang}.

 \section{Discussion}

We have determined the time dependence of the electron spin relaxation rate arising from its interaction with nuclear spins for an ensemble of QDs, or equivalently,  averaged over a large number of successive measurements of a single dot. This gives
us the value of $T_2^*$ usually measured in ensemble experiments.

The electron spin relaxation time was studied in three types of QDs: GaAs natural quantum dots, which are  island-like dots formed by a fluctuation of the GaAs quantum well thickness \cite{Gammon}; self organized InAs QDs \cite{Petroff}, and nanocrystal CdSe QDs \cite{Gupta}. We can estimate the value of $T_\Delta$ for each of these dots.  In GaAs natural QDs, using values of $A^j$ from Ref.\cite{Paget} and Eq.(6), and taking the dot to contain $10^5$ nuclei, Eq.(11) gives $T_\Delta\sim 1.0$\,ns
  The values of $A^j$ are not experimentally determined for InAs. Assuming that compound ionicity does not significantly vary among the semiconductors GaAs, InSb and InAs\cite{Paget,Phillips},  we take the hyperfine constants, $A^j$, for As and In ions from \cite{Paget} and \cite{guron}: $A^j_{As}=47\mu$eV and $A^j_{In}=56\mu$eV,  respectively.  The large value of the In nuclear spin, $I=9/2$, strongly effects the magnitude of the hyperfine interaction in InAs, and shortens the spin dephasing time. To estimate this time we need to know the number of nuclei in the electron localization volume. In the self organized QDs, where the electron wave function strongly depends on the QD shape and barrier height, we  used the geometrical volume of the dot to determine the number of nuclei, $N_L$, in Eq.(7).  The QDs studied in \cite{Petroff} contain 1000-4000  nuclei \cite{privite}.  This leads to  electron spin relaxation times  on the order of 50-100\,ps, which is close to the value of $gT_2^*$ measured in these dots.
The   hyperfine interaction constants in CdSe  are also not experimentally determined. In this material only a fraction of the nuclei (25 \% of the Cd ions) have a magnetic moment and these have spin $I=1/2$.  As a result, the electron spin interaction with the nuclei in these QDs is weak. We shall take for the  hyperfine constant in CdSe  half that of  $Cd^{111}$ \cite{koh}, assuming completely covalent bonding in this material. This leads to  $A^j_{Cd}=12\mu$eV.  Taking the CdSe lattice constant $a_0\approx 4.2$\AA\ \cite{heterostructures} we calculate $N_L=3,500$ for the  ground 1S electron state in spherical nanocrystals with a  radius 28\AA.  Only 1/8 of these nuclei contribute to dephasing. This leads to $T_\Delta\approx 1.6$\,ns, which is consistent with the experimental measurement $T_2^*=2.5$\,ns \cite{Gupta}

In conclusion,  we developed a theory of electron spin relaxation in QDs
 arising from their hyperfine interaction with the QD nuclei. The relaxation is determined by three physical processes: (i) -- the precession of the electron spin
in the hyperfine field of the frozen fluctuation of the nuclear spins; (ii) -- the precession of the nuclear spins in the hyperfine field of the electron; and (iii) -- the precession of a nuclear spin in the dipole field of its neighbors.
These processes have three disparate characteristic times. For GaAs QDs with $10^5$ nuclei, they are $\sim 1$\,ns, $\sim 300$\,ns, and $\sim 100\mu$s, respectively. The last of these times is so long that many other electron spin relaxation mechanisms can be more important on this time scale. An external magnetic field suppresses the relaxation of the spin component along the magnetic field. The transverse components of the electron spin polarization relax completely in a time on the order of the electron precession period in the field of the frozen nuclear fluctuation. Comparison with experimental data shows that the hyperfine interaction with nuclei  is the dominant mechanism of electron spin relaxation  in QDs.

{\it Acknowledgments.} All  the authors thank the DARPA/Spin program for financial support,  Al.L.E. and M.R. were supported by the Office of Naval Research,
 and I. A. Merkulov received support from CRDF and RFBI grants.

\section*{Appendix: The Number of nuclei interacting with a localized electron in a QD}

 The number of nuclei, $N$,  "interacting"  with a localized electron and forming an  electron-nuclear system in thermodynamic equilibrium is generally a time dependent quantity. Formally, the electron interacts at all times with all those nuclei  of the crystal lattice
sites where its wave function is nonzero. However, the strength of this interaction   decreases with the distance from the QD center as the square of the electron wave function, and the random magnetic field $\mbox{\boldmath $B$}_N$, which determines the electron spin precession is determined only by the relatively small number of nuclei, $N_L$, in the"effective" volume of the electron localization $V_L$ (see Eq.8). This volume (or alternatively $N_L$)  completely determines the electron spin relaxation
at short times $t< T_N$ during which nuclear spins, and, therefore, the corresponding field $\mbox{\boldmath $B$}_N$ do not change their directions.

The precession of the nuclear spins in the hyperfine field of the electron causes time dependent changes in $\mbox{\boldmath $B$}_N$. This time dependent $\mbox{\boldmath $B$}_N(t)$ in each dot is determined by a macroscopic number of initial conditions. The correlation $\langle\mbox{\boldmath $B$}_N(t)$$\mbox{\boldmath $B$}_N(0)\rangle$ (see Eq.14) 
is now an  important characteristic  of an ensemble of quantum dots. The ensemble of localized electrons completely loses  its memory of the initial spin polarization if this correlation is zero. Although $\langle\mbox{\boldmath $B$}_N(t)$$\mbox{\boldmath $B$}_N(0)\rangle$ can be calculated for any time, $t$, in principle, in the long time limit
 $t\gg T_N$, when the values of $\mbox{\boldmath $B$}_N(t)$ are randomly distributed, one can find this correlation  using statistical physics techniques.

Let us separate the nuclei interacting with a localized electron during a time $t$ into two groups: (1) those nuclei 
whose spin rotates through an angle $\omega_it\geq 1$ and (2) those whose spin rotates through an  angle
$\omega_it\leq 1$, where $\omega_i$ is the precession frequency of the nuclear spin in the hyperfine field of the localized electron.  The second group can be considered as nuclei that "do not interact" with the electron. For times
$t\gg T_N$ the first group automatically includes the $N_L$ nuclei that determines the field $\mbox{\boldmath $B$}_N$. The number of nuclei $N(t)$ "interacting" with the electron, and thus belong to the first group, increases with time and determines $\mbox{\boldmath $B$}_N(t)$, whose  variation is restricted by 
conservation of the total spin of the electron-nuclear spin system.  Randomness of the distribution  of the $N(t)$ nuclear spins limited by the conservation of $B_N$ and $\mbox{\boldmath $I$}_\Sigma$ is the one major assumption of this statistical model.

For nuclei that interact strongly through the "indirect" hyperfine interaction, i.e.
nuclei whose spins rotate through an  angle $\omega_it\gg 1$ in the electron hyperfine field, 
this assumption is justified.
 The error in the statistical approach  is connected with the border nuclei for which $\omega_it\sim 1$.  These nuclei strongly interact with
the electron, but, on the other hand, they still  "remember" their initial direction at $t=0$. The number of such nuclei is on the order of $\Delta N\approx t\partial N/\partial t$.  As a  result the relative error of our estimate is
$\delta N=\Delta N(t)/N(t)\approx (t/N)\partial N/\partial t$.

The time dependence of $N(t)/N_L$ and $\delta N$ in spherical QDs with an infinite barrier is shown in Fig.4. The time is measured in the unit of $T_N^0\sim (32/\pi)(\hbar/A)(a^3/v_0)$, the  nuclear precession time at the center of the QD, where $a$ is the QD radius. In GaAs QDs with $10^5$ nuclei, this time is on the order of microseconds. One sees that $N<N_L$ for $t<10T_N^0$.  In this time period
$\delta N\ge 1$ and our statistical approach is not valid. However $N(t)$ increases and $\delta N(t)$ decreases rapidly with time, and  $\delta N\sim 0.25$ at $t\approx 20T_N^0$. After this time the nuclei which give the main contribution to the magnetic field, $\mbox{\boldmath B}_N$, acting on the electron
 are in statistical equilibrium. This allows us to use Eq.(\ref{gamma}).

\begin{figure} [p]
\caption{The dependence of the ensemble averaged electron spin polarization on the ratio of dispersion of the electron hyperfine field  acting on the nuclei to its ensemble averaged value.}
\label{F1}
\end{figure}

 \begin{figure}[p]
 \caption{The time dependence of the longitudinal, ($a$), and transverse,  ($b$) and ($c$), components of the ensemble averaged electron spin polarization calculated for different magnetic fields. The curves are calculated for $\beta=0,1,2,3,5$ respectively.}
\label{F2}
\end{figure}

\begin{figure}[p]
 \caption{The magnetic field dependence of the longitudinal, ($a$), and transverse, ($b$) and ($c$), components of the steady state electron spin polarization. Calculations are done for electron lifetimes $\tau/T_\Delta =0,~0.2,~0.4,~0.8,~1.4,~3.5,~{\rm and}~10$.}
\label{F3}
\end{figure}

\begin{figure}[p]
\caption{Time dependence of the ratio of the number of nuclei, $N(t)$, "interacting" with an electron, in a spherical QD with an infinite potential barrier, to $N_L$ and the relative error, $\delta N(t)$.}
\label{F4}
\end{figure}


\begin{references}
\bibitem{diffusion} Spin  conservation can be destroyed  by nuclear dipole-dipole  interaction  effects or by spin diffusion.
\end{references}
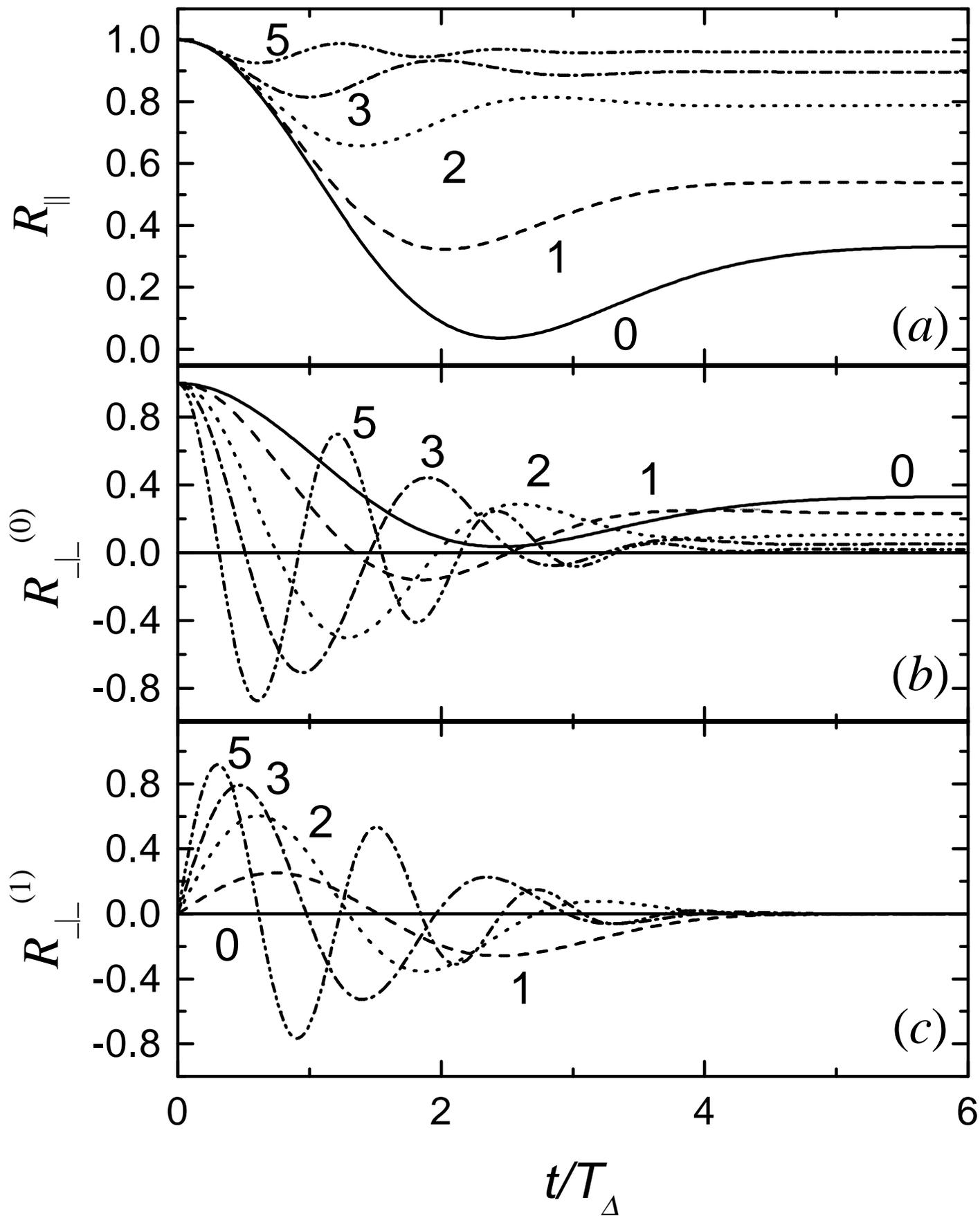
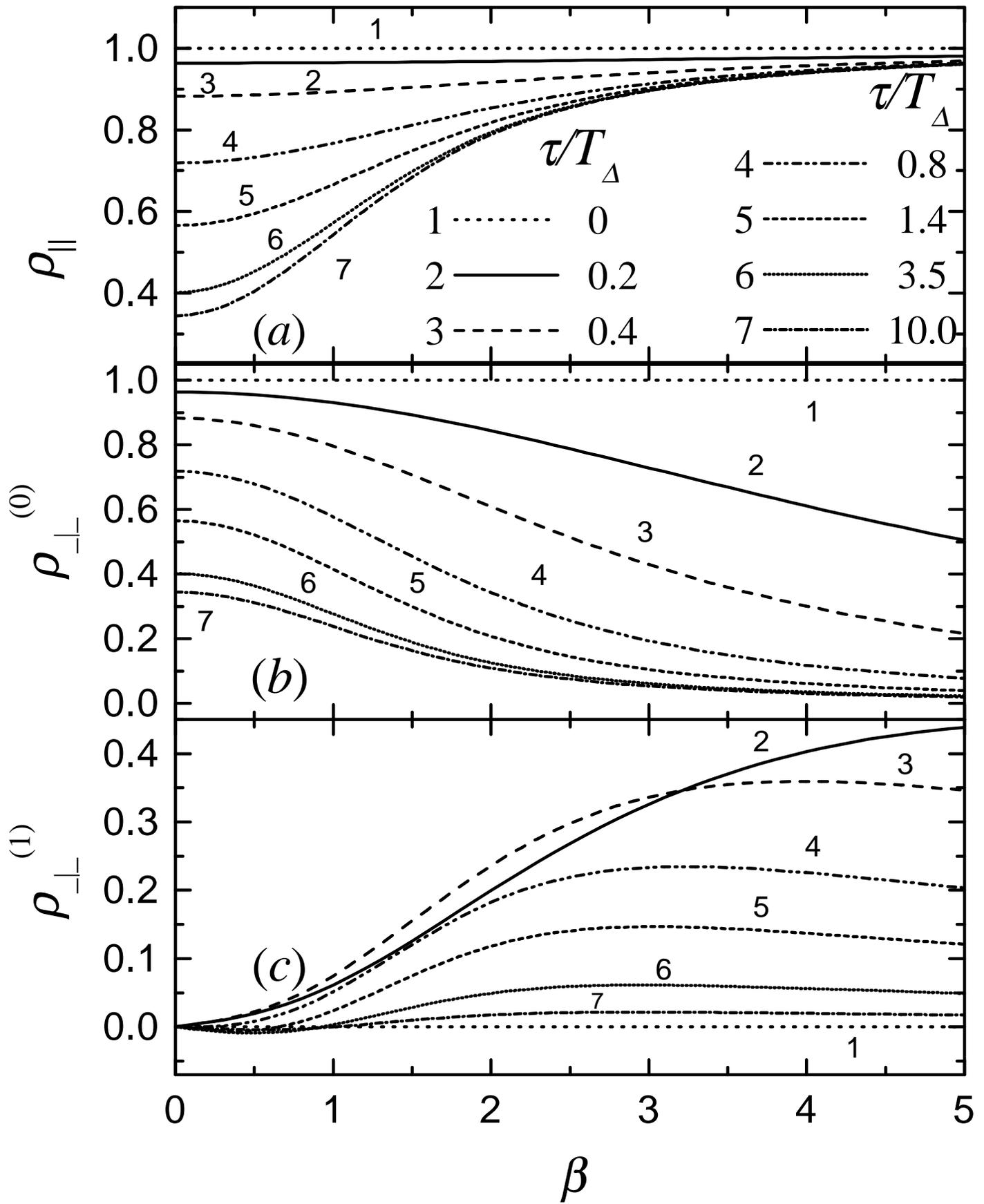
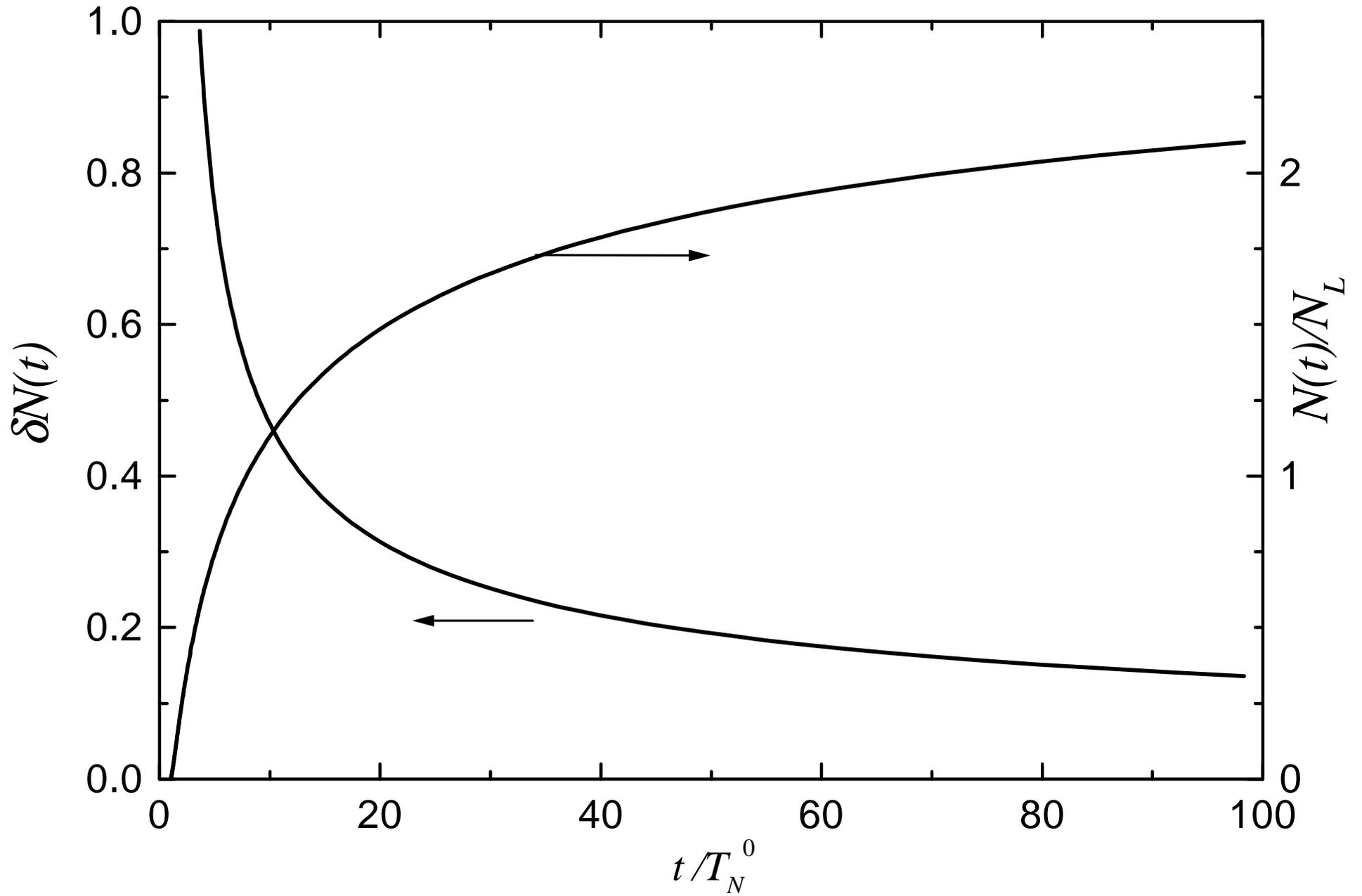
\end{document}